%% file: main.tex
\documentclass{fsttcs}

\usepackage{amsmath,amssymb,latexsym,stmaryrd}
\usepackage[algoruled,vlined,english]{algorithm2e}
\usepackage[normalem]{ulem}
\usepackage{setspace}
\usepackage[backgroundcolor=orange!50!white]{todonotes}

\def\@envspa{\hspace{0.3em}}
\def\@sa{\hspace{-0.2em}}
\def\@sb{\hspace{0.5em}}
\def\@sc{\hspace{-0.1em}}

\makeatletter

\begingroup \catcode `|=0 \catcode `[= 1
\catcode`]=2 \catcode `\{=12 \catcode `\}=12
\catcode`\\=12 |gdef|@xcomment#1\end{comment}[|end[comment]]
|endgroup
\def\@comment{\let\do\@makeother \dospecials\catcode`\^^M=10\def\par{}}
\def\begincomment{\@comment\@xcomment}
\makeatother

\def\set#1{{\left\{ #1 \right\}}}
\def\tuple#1{{\left\langle #1 \right\rangle}}
\def\multi#1{{\llbracket #1 \rrbracket}}
\def\nats{{\mathbb{N}}}

\newcommand{\multiset}[1]{{\mathbb{M}[ #1 ]}}

\def\mmap{\mathbf{m}}

\newcommand{\fire}[1]{\left[ {#1}\right\rangle}
\def\prod{\mathcal{P}}
\def\cfl{\mathsf{CFL}}
\def\cfg{\mathsf{CFG}}
\def\pn{\mathsf{PN}}
\def\pni{\mathsf{PNI}}
\def\pnl{\mathsf{PNL}}
\def\pnw{\mathsf{PNW}}
\def\pnwl{\mathsf{PNWL}}

\def\addto{\mathit{add\_\mathord{\ast}\_to}}
\def\subto{\mathit{sub\_\mathord{\ast}\_to}}

\begin{document}
\title{Approximating Petri Net Reachability Along Context-free Traces}
\runningtitle{Approximating Petri Net Reachability Along Context-free Traces}

\author{Mohamed Faouzi Atig\inst{1}, Pierre Ganty\inst{2}}

\institute{1}{Uppsala University, Sweden}
\institute{2}{IMDEA Software Institute, Spain}

\runningauthors{M. F. Atig, P. Ganty}

\begin{abstract}
We investigate the problem asking whether the intersection of a
context-free language (\(\cfl\)) and a Petri net language (\(\pnl\)) is empty.
Our contribution to solve this long-standing problem which relates, for
instance, to the reachability analysis of recursive programs over unbounded
data domain, is to identify a class of \(\cfl\)s called the finite-index
\(\cfl\)s for which the problem is decidable.  The \(k\)-index approximation of
a \(\cfl\)  can be obtained by discarding all the words that cannot be derived
within a budget \(k\) on the number of occurrences of non-terminals.  A
finite-index \(\cfl\) is thus a \(\cfl\) which coincides with its \(k\)-index
approximation for some \(k\). We decide whether the intersection of a finite-index \(\cfl\) and a \(\pnl\) is empty by reducing it to the reachability
problem of Petri nets with {\em weak} inhibitor arcs, a class of
systems with infinitely many states for which reachability is known to be
decidable.  Conversely, we show that the reachability problem for a Petri net
with weak inhibitor arcs reduces to the emptiness problem of a finite-index
\(\cfl\) intersected with a \(\pnl\).
\end{abstract}

\section{Introduction}


Automated verification of infinite-state systems, for instance programs with
(recursive) procedures and integer variables, is an important and a highly
challenging problem.  Pushdown automata (or equivalently context-free grammars) have been
proposed as an adequate formalism to model procedural programs. However
pushdown automata require finiteness of the data domain which is typically
obtained by abstracting the program's data, for instance, using the predicate
abstraction techniques \cite{Cousot77-POPL,GS97}.  In many cases, reasoning
over finite abstract domains yields to a too coarse analysis and is therefore
not precise. To palliate this problem, it is natural to model a procedural
program with integer variables as a pushdown automaton manipulating counters.
In general, pushdown automata with counters are Turing powerful 
which implies that basic decision problems are undecidable (this is true even
for the case finite-state automata with counters).

Therefore one has to look for restrictions on the model which retain sufficient
expressiveness while allowing basic properties like reachability to be
algorithmically verified. One such restriction is to forbid the test of a
counter and a constant for equality.  In fact, forbidding test for equality
implies the decidability of the reachability problem  for the case of finite-state
automata with counters (i.e. Petri nets \cite{LEROUX-POPL2011,Reinhardt}). 

The verification problem  for pushdown automata with (restricted) counters
boils down to check whether a context-free language (\(\cfl\)) and a Petri net
language (\(\pnl\)) are disjoint or not. We denote this last problem
\(\pnl\cap\cfl\stackrel{?}{=}\emptyset\).

		

The decidability of \(\pnl\cap\cfl\stackrel{?}{=}\emptyset\) is open and lies at the very edge of our
comprehension of infinite-state systems.  We see two breakthroughs contributing
to this question.  First, determining the emptiness of a \(\pnl\) was known to
be decidable as early as the eighties. Then, in 2006, Reinhardt
\cite{Reinhardt} lifted this result to an extension of \(\pn\) with inhibitor
arcs (that allow to test if a counter equals 0) {which must satisfy some
additional topological conditions. By imposing a topology on the tests for
zero,} Reinhardt prevents his model to acquire Turing powerful capabilities. We
call his model \(\pnw\) and the languages thereof \(\pnwl\).


Our contribution to the decidability of \(\pnl\cap\cfl\stackrel{?}{=}\emptyset\) comes under the form
of a partial answer which is better understood in terms of underapproximation.
In fact, given a \(\pnl\) \(L_1\) and the language \(L\) of a context-free
grammar we replace \(L\) by a subset \(L'\) which is obtained by discarding
from \(L\) all the words that cannot be derived within a given budget $k \in \mathbb{N}$ on the
number of non-terminal symbols. (In fact, the subset $L'$  contains any word of $L$ that can be generated  by a  derivation  that  contains at most $k$ non-terminal symbols at each derivation step.)  We show how to compute \(L'\) by annotating the
variables of the context-free grammar for \(L\) with an allowance. What is
particularly appealing is that the coverage of \(L\) increases with the allowance.
{%
Approximations induced by allowances are non-trivial: every regular or
linear language is captured exactly with an allowance of \(1\), \(L'\) coincides with \(L\)
when the allowance is unbounded, and under commutativity of concatenation
\(L'\) coincides with \(L\) for some allowance
\(k\in\nats\).}

We call finite-index \(\cfl\), or \(\text{fi}\cfl\) for short, a context-free
language where each of its words can be derived within a given budget.  In this
paper, we prove the decidability of
\(\pnl\cap\text{fi}\cfl\stackrel{?}{=}\emptyset\) by reducing it to the
emptiness problem of \(\pnwl\). We also prove the converse reduction; showing
those two problems are equivalent. Hence, we offer a whole new perspective on
the emptiness problem for \(\pnwl\) and \(\pnl\cap\cfl\).  

To conclude the introduction let us mention the recent result of \cite{RB-mfcs11} which builds on
\cite{LEROUX-POPL2011} to give an alternative proof of Reinhardt's result
(\(\pnw\) reachability is decidable) for the particular case where one counter
only can be tested for zero.

\input{pre.tex}
\input{pncfl-ipn.tex}
\input{part2nprg.tex}
\input{pnw-pncfl.tex}

\section{Conclusion}
In this paper, we have defined  the class finite-index context-free languages  (which is an interesting sub-class of  context-free languages). We have shown that  the problem of checking whether the intersection of a finite-index context-free language and a Petri net language is empty is decidable. This result is obtained through a non-trivial reduction to  the reachability problem for Petri nets with weak inhibitor arcs. On the other hand, we have proved that the reachability problem for Petri nets with weak inhibitor arcs  can be reduced to the the emptiness problem of the language obtained from the intersection of a finite-index context-free language and a Petri net language, which implies by \cite{Lipton} that the latter is EXPSPACE-hard.

\begin{spacing}{0.9}
\bibliographystyle{abbrv}
\bibliography{ref}
\end{spacing}

\newpage
\appendix

\section{Missing Net programs}
Alg.~\ref{alg:subto} gives the net program which implements the call \(\subto(\mathbf{M_{\sf i}}[\ell],\mathbf{M_{\sf f}}[\ell])\).

\begin{algorithm}[H]
\SetKw{goto}{goto}
\SetKw{gosub}{gosub}
\SetKw{Kor}{or}
\SetKw{halt}{halt}
	\nlset{\(\mathbf{sub\_to}_{\ell}\)}\goto \textbf{exit} \Kor \(s_1\) \Kor \ldots \Kor \(s_d\) \;
	\nlset{\(s_1\):}\(\mathbf{M}_{\sf i}[\ell][1]:=\mathbf{M}_{\sf i}[\ell][1]-1\)\;
	\(\mathbf{M}_{\sf f}[\ell][1]:=\mathbf{M}_{\sf f}[\ell][1]-1\)\;
	\goto \textbf{\(\mathbf{sub\_to}_{\ell}\)}\;
	[\ldots]\;
	\nlset{\(s_d\):}\(\mathbf{M}_{\sf i}[\ell][d]:=\mathbf{M}_{\sf i}[\ell][d]-1\)\;
	\(\mathbf{M}_{\sf f}[\ell][d]:=\mathbf{M}_{\sf f}[\ell][d]-1\)\;
	\goto \(\mathbf{sub\_to}_{\ell}\)\;
	\nlset{\textbf{exit:}} \Return{}\;
	\caption{\label{alg:subto}}
\end{algorithm}
Alg.~\ref{alg:addto} implements the call \(\addto(\mathbf{M_{\sf i}}[\ell],\mathbf{M_{\sf f}}[\ell-1])\).

\begin{algorithm}[H]
\SetKw{goto}{goto}
\SetKw{gosub}{gosub}
\SetKw{Kor}{or}
\SetKw{halt}{halt}
\nlset{\(\mathbf{add\_to\_i}_{\ell}\_\mathbf{f}_{\ell-1}\)}\goto \textbf{exit} \Kor \(s_1\) \Kor \ldots \Kor \(s_d\) \;
	\nlset{\(s_1\):}\(\mathbf{M}_{\sf i}[\ell][1]:=\mathbf{M}_{\sf i}[\ell][1]+1\)\;
	\(\mathbf{M}_{\sf f}[\ell-1][1]:=\mathbf{M}_{\sf f}[\ell-1][1]+1\)\;
	\goto \(\mathbf{add\_to\_i}_{\ell}\_\mathbf{f}_{\ell-1}\)\;
	[\ldots]\;
	\nlset{\(s_d\):}\(\mathbf{M}_{\sf i}[\ell][d]:=\mathbf{M}_{\sf i}[\ell][d]+1\)\;
	\(\mathbf{M}_{\sf f}[\ell-1][d]:=\mathbf{M}_{\sf f}[\ell-1][d]+1\)\;
	\goto \(\mathbf{add\_to\_i}_{\ell}\_\mathbf{f}_{\ell-1}\)\;
	\nlset{\textbf{exit:}} \Return{}\;
	\caption{\label{alg:addto}}
\end{algorithm}
Alg.~\ref{alg:trfromto} implements the call \(\mathit{transfer\_from\_to}(\mathbf{M_{\sf f}}[\ell],\mathbf{M_{\sf f}}[\ell-1])\).

\begin{algorithm}[H]
\SetKw{goto}{goto}
\SetKw{gosub}{gosub}
\SetKw{Kor}{or}
\SetKw{halt}{halt}
\nlset{\(\mathbf{tr\_f}_{\ell}\mathbf{\_f}_{\ell-1}\)}\goto \textbf{exit} \Kor \(s_1\) \Kor \ldots \Kor \(s_d\) \;
	\nlset{\(s_1\):} \(\mathbf{M}_{\sf f}[\ell][1]:=\mathbf{M}_{\sf f}[\ell][1]-1\)\;
	\(\mathbf{M}_{\sf f}[\ell-1][1]:=\mathbf{M}_{\sf f}[\ell-1][1]+1\)\;
	\goto \(\mathbf{tr\_f}_{\ell}\mathbf{\_f}_{\ell-1}\)\;
	[\ldots]\;
	\nlset{\(s_d\):}\(\mathbf{M}_{\sf f}[\ell][d]:=\mathbf{M}_{\sf f}[\ell][d]-1\)\;
	\(\mathbf{M}_{\sf f}[\ell-1][d]:=\mathbf{M}_{\sf f}[\ell-1][d]+1\)\;
	\goto \(\mathbf{tr\_f}_{\ell}\mathbf{\_f}_{\ell-1}\)\;
	\nlset{\textbf{exit:}} \Return{}\;
	\caption{\label{alg:trfromto}}
\end{algorithm}

\pagebreak
\section{Missing Proofs}

\subsection{Proof of Lemma~\ref{lem1}}
\begin{proof}
	Let \(w\in\Sigma^*\), we shall demonstrate that
	\(A^{[k]}\Rightarrow^* w\) if{}f there exists a derivation \(A\Rightarrow^* w\)
	that is \(k+1\) index bounded.
	
\noindent %
{\bf Only if.}
We have \(A^{[k]}\Rightarrow^{\ell} w\) for some
\(\ell\in\nats\setminus\set{0}\). The proof is done by induction on \(\ell\).
For the case \(\ell=1\), we have \(A^{[k]}\Rightarrow w\), hence that \(
(A^{[k]},w)\in\prod^{[k]} \) and \( (A,w)\in\prod\) by definition of
\(G^{[k]}\) and finally that \(A\Rightarrow w\) is \(1\leq k+1\) index bounded.
For the case \(\ell>1\), the definition of \(G^{[k]}\) shows that there exists
a derivation of the form (1) \(A^{[k]}\Rightarrow B^{[k-1]} C^{[k]}
\Rightarrow^i w_1 C^{[k]} \Rightarrow^j w_1w_2=w \) where \(i+j=\ell-1\) or (2)
\(A^{[k]}\Rightarrow B^{[k]} C^{[k-1]} \Rightarrow^{j'} B^{[k]} w_2 \Rightarrow^{i'} w_1w_2=w \)
where \(i'+j'=\ell-1\) which is treated similarly.  
Assume case (1) holds. Because \(B^{[k-1]}\Rightarrow^i w_1\) where \(i<\ell\)
we find, by induction hypothesis, that there exists a derivation \(B\Rightarrow^*
w_1\) that is \(k\) index bounded. Also, since \(C^{[k]}\Rightarrow^j w_2\) where \(j<\ell\), 
the induction hypothesis shows that there exists a derivation \(C\Rightarrow^*
w_2\) that is \(k+1\) index bounded.  Finally,
we conclude from  \((A^{[k]},B^{[k-1]} C^{[k]})\in\prod^{[k]}\), that \((A,BC)\in\prod\), hence
that there exists a derivation \(A\Rightarrow BC\Rightarrow^* w_1
C\Rightarrow^* w_1 w_2 =w\) that is \(k+1\) index bounded and we are done.

\noindent %
{\bf If.}
Let \(A\Rightarrow^{\ell} w\) for some
\(\ell\in\nats\setminus\set{0}\) be a \(k+1\) index bounded derivation. 
The proof is done by induction on \(\ell\).
For the case \(\ell=1\), we conclude from \(A\Rightarrow w\) is \(k+1\) index bounded that
\((A,w)\in\prod\) by definition of \(G\), hence that \(
(A^{[k]},w)\in\prod^{[k]} \) by definition of \(G^{[k]}\) and finally that
\(A^{[k]}\Rightarrow w\).

For the case \(\ell>1\), there is a \(k+1\) index bounded derivation of the
form \(A\Rightarrow BC\Rightarrow^{\ell-1} w\) such that one of the following
derivation is \(k+1\) index bounded: \(A\Rightarrow BC\Rightarrow^i w_1
C\Rightarrow^j w_1 w_2=w\) or \(A\Rightarrow BC\Rightarrow^j B w_2\Rightarrow^i
w_1 w_2=w\) where \(i+j=\ell-1\).

Assume the former case holds (the other is handled similarly).  Since the
derivation is \(k+1\) index bounded we find that \(B\Rightarrow^i w_1\) is
\(k\) index bounded and \(C\Rightarrow^{j} w_2\) is \(k+1\) bounded.  Because
\(i<\ell\) and \(j<\ell\) we find, by induction hypothesis, that \(w_1\in L(B^{[k-1]})\) and \(w_2\in L(C^{[k]})\).
Finally, \(A\Rightarrow BC\) shows that \( (A,BC)\in\prod \), 
hence we deduce that \( \set{(A^{[k]}, B^{[k-1]}C^{[k]}),(A^{[k]},
B^{[k]}C^{[k-1]})}\subseteq \prod^{[k]}\), and finally that \(A^{[k]}\Rightarrow^* w\) holds.\qed
\end{proof}


\subsection{Proof of Lemma~\ref{sec5.lem7}}

\begin{proof} 
The proof is done by induction on $\ell$.

\noindent {\bf Basis.} $\ell=0$. 
Let \(w\in L_0\), that is \(w\in T_0^{k}\) for some \(k\in\nats\).
The proof is by induction on \(k\). The case \(k=0\) (\(w=\varepsilon\)) is trivially
solved. Let \(k>0\), then \(w\) can be decomposed in \(w_1,\ldots,w_k\)
where each \(w_i\in T_0\) for \(i\in\set{1,\ldots,k}\) and
\(w_i\) is necesarily of the form \(v_j \cdot t_j \cdot u_j\). 
Finally since the firing of $v_j \cdot t_j \cdot u_j\in T_0$ keeps unchanged the total number of
tokens in \(\set{s_i,r_i}\) for each \(i\in\set{1,\ldots,n}\) then
so does all \(w\in T_0\) and we are done.

\noindent {\bf Step.} $\ell>0$.   
The definition of \(L_{\ell}\) shows that \(w\in((\tuple{p_{\ell},c_{\ell}}\star
L_{\ell-1}) \cup T_{\ell})^k\) for some \(k\in\nats\). The proof is done by
induction on \(k\). The case \(k=0\) (\(w=\epsilon\)) is trivially solved. For
\(k>0\) we have that \(w=w_1\cdots w_k\) where \(w_i\in
\tuple{p_{\ell},c_{\ell}}\star L_{\ell-1}\) or \(w_i\in T_{\ell}\).  If
\(w_1\in T_{\ell}\), then using the above reasoning we find that the the firing
of any \(w\in T_{\ell}\) keeps unchanged the total number of tokens in
\(\set{s_i,r_i}\) for each \(i\in\set{1,\ldots,n}\).  If
\(w_1\in \tuple{p_{\ell},c_{\ell}}\star L_{\ell-1}\) then \(w_1= p_{\ell}^{i} v
c_{\ell}^{i}\) for some \(i\in\nats\), \(v\in L_{\ell-1}\).
Since the result holds for every \(v\in L_{\ell-1}\) by induction hypothesis,
we find that it also holds for \(w_1\) by definition of \(p_{\ell}\) and
\(c_{\ell}\) and because they fire an equal number of times.  Finally we use
the induction hypothesis on \(w_2\cdots w_k\) (we can because \(w_2\cdots
w_k\in L_{\ell}\)) and we are done.\qed
\end{proof}

\subsection{Proof of Lemma~\ref{sec5.lem8}}

\begin{proof} The proof is done by induction on \(\ell\). 
\medskip

\noindent
{\bf Basis.} $\ell=0$.
\(w\in L_{0}=T_0^*\) and every transition \(t\) occurring in \(w|_{T}\) is such
that \(Z(t)=\emptyset\), hence the def.\ of \(N'\) and \(\mmap_a\fire{w}_{N'}\mmap_b\) show that \(S(\mmap_{a})\fire{w|_{T}}_{N} S(\mmap_b)\).

\medskip

\noindent {\bf Step.} $\ell>0$.
The definition of \(L_{\ell}\) shows that \(w\in((\tuple{p_{\ell},c_{\ell}}\star
L_{\ell-1}) \cup T_{\ell})^k\) for some \(k\in\nats\). The proof is done by
induction on \(k\). The case \(k=0\) (\(w=\epsilon\)) is trivially solved. For
\(k>0\) we have that \(w=w_1\cdots w_k\) where \(w_i\in
\tuple{p_{\ell},c_{\ell}}\star L_{\ell-1}\) or \(w_i\in T_{\ell}\).  If
\(w_1\in \tuple{p_{\ell},c_{\ell}}\star L_{\ell-1}\) then \(w_1= p_{\ell}^{i} v
c_{\ell}^{i}\) for some \(i\in\nats\), \(v\in L_{\ell-1}\).  Let
\(\mmap_0,\mmap'_{0},\mmap'_{1},\mmap_{1}\) such that
\(\mmap_a=\mmap_0\fire{p_{\ell}^{i}}\mmap'_{0}\fire{v}\mmap'_{1}\fire{c_{\ell}^{i}}\mmap_{1}\).
We conclude from \((S_{\ell}\cup R_{\ell})(\mmap_{a})=\varnothing\) and
\(p_{\ell}^{i}\) that \((S_{\ell-1}\cup
R_{\ell-1})(\mmap'_{0})=\varnothing\).  Next Lem.~\ref{sec5.lem7} shows that \(
(S_{\ell-1}\cup R_{\ell-1})(\mmap'_{1})=\varnothing\).  Hence, the induction
hypothesis on \(L_{\ell-1}\) shows that \(S(\mmap'_{0})\fire{v|_{T}}_N
S(\mmap'_{1})\). Finally the definition of \(w_1\) shows that \(w_1|_{T}=v|_{T}\),
hence that \(S(\mmap'_{0})\fire{w_1|_{T}}_N S(\mmap'_{1})\), and finally that
\(S(\mmap_0)\fire{w_1|_{T}}_N S(\mmap_{1})\) since \(S(\mmap_0)=S(\mmap'_{0})\) and
\(S(\mmap_{1})=S(\mmap'_{1})\).  
Also from the assumption \( (S_{\ell}\cup R_{\ell})(\mmap_0)=\varnothing\),
\(w_1\in L_{\ell}\) and Lem.~\ref{sec5.lem7} we conclude that \( (S_{\ell}\cup
R_{\ell})(\mmap_{1})=\varnothing\).

Let us now turn to the case \(w_1\in T_{\ell}\).  Let \(\mmap_1\) such that
\(\mmap_a\fire{w_1}\mmap_1\), we conclude from \((S_{\ell}\cup
R_{\ell})(\mmap_{a})=\varnothing\), \(w_1\in L_{\ell}\) and Lem.~\ref{sec5.lem7} that
\((S_{\ell}\cup R_{\ell})(\mmap_1)=\varnothing\), hence that
\(S(\mmap_{a})\fire{w_1|_{T}}_{N}S(\mmap_1)\) since \(w_1|_{T}=t_j\),
\(Z(t_j)=S_{\ell}\) and \(S_{\ell}(\mmap_{a})=\varnothing\). 

Finally we use the induction hypothesis on \(w_2\cdots w_k\) (we can because
(1) \(w_2\cdots w_k\in  L_{\ell}\) and (2) we have shown that \((S_{\ell}\cup
R_{\ell})(\mmap_1)=\varnothing\) in both cases) and we are done.
\qed
\end{proof}

\subsection{Proof of Lemma~\ref{sec5.lem9}}

\begin{proof} The proof is done by induction on $\ell$.

\medskip

\noindent
{\bf Basis.} $\ell=0$. First, let us observe that, since \(\ell=0\), the predicates
\(S_{\ell}(\mu_1)=S_{\ell}(\mu_2)=\varnothing\) and \(R_{\ell}(\mmap_1)=R_{\ell}(\mmap_2)=\varnothing\)
are vacuously true. Let $\mu_1 \fire{u}_N \mu_2$ where $u \in {L_0|_T}$. Then, there is a word $w \in L_0$ such
that $u=w|_T$. Let $\mmap_1\in\multiset{S'}$ defined as follows:
$S(\mmap_1)=\mu_1$, and $\mmap_1(r_i)=|w|$ for all $i \in \set{1,\ldots,n}$.
Then, we have $\mmap_1\fire{w}_{N'}$ which yields $\mmap_2$ since there are enough
tokens in the places $R_{n}$. Moreover, we have $S(\mmap_2)=\mu_2$ since no
transition in $\set{p_1,c_1,\ldots,p_n,c_n}$ has an arc to a place in $S$.

\medskip

\noindent {\bf Step.} $\ell>0$.  Since there is $u \in
{L_\ell|_T}$ such that $\mu_1 \fire{u}_{N} \mu_2$, then either case must hold: 

\begin{itemize}
\item {\bf Case 1:} $u \in L_{\ell-1}|_T$. Then, we can use the induction
  hypothesis to show that there are $\mmap'_1, \mmap'_2 \in \multiset{S'}$ and
  $w' \in L_{\ell-1}$  such that $S(\mmap'_1)=\mu_1$, $S(\mmap'_2)=\mu_2$,
  $R_{\ell-1}(\mmap'_1)=R_{\ell-1}(\mmap'_2)=\varnothing$, and
  $\mmap'_1\fire{w'}_{N'}\mmap'_{2}$. 
  Next, Lem.~\ref{sec5.lem7} shows that
  \(\mmap'_1(s_{\ell})+\mmap'_1(r_{\ell})=\mmap'_2(s_{\ell})+\mmap'_2(r_{\ell})\),
  hence that \(\mmap'_1(r_{\ell})=\mmap'_2(r_{\ell})\) since
  \(S_{\ell}(\mu_i)=\varnothing\) and \(S(\mmap'_{i})=\mu_i\) for
  \(i\in\set{1,2}\). 
  Let  $w= p_{\ell}^{\, j}\,  w' \,
  c_{\ell}^{\, j}\in L_{\ell}$ where $j= \mmap'_1(r_\ell)$,
  and let \(\mmap_1, \mmap_2\in\multiset{S'}\) such that $(S'\setminus\set{r_{\ell}})(\mmap_i)=(S'\setminus\set{r_{\ell}})(\mmap'_i)$ and \(\mmap_i(r_{\ell})=0\) for \(i\in\set{1,2}\).
  From the above we find that (i) \(S(\mmap_i)=S(\mmap'_i)=\mu_i\)  for
  \(i\in\set{1,2}\), (ii) \(R_{\ell}(\mmap_1)=R_{\ell}(\mmap_2)=\varnothing\) (since \(R_{\ell-1}(\mmap'_1)=R_{\ell-1}(\mmap'_2)=\varnothing\) and by def.\ of \(\mmap_1,\mmap_2\)) and
  (iii) \(\mmap_1\fire{w}_{N'}\mmap_2\) (since \(\mmap'_1(r_{\ell})=\mmap'_2(r_{\ell})\)
  we can show that \(\mmap_1\fire{p_{\ell}^{j}}\mmap'_1\fire{w'}\mmap'_2\fire{c_{\ell}^{j}}\mmap_2\)) and we are done.
\item {\bf Case 2:} $u=w_0 t_{i_1} w_1 t_{i_2} w_2 \cdots t_{i_{k}} w_k$ for
  some  $w_1,\ldots,w_k  \in L_{\ell-1}|_T$ and $t_{i_1}, \ldots, t_{i_k}\in T_{\ell}|_T$ 
  (also $Z(t_{i_1})=\cdots=Z(t_{i_k})=S_{\ell}$).
  To simplify the presentation, we assume  that $k=1$. (The general case
  can be handled in the same way.)  Then, there are $\mu'_1, \mu'_2 \in
  \multiset{S}$ such that $\mu_1  \fire{w_0}  \mu'_1  \fire{t_{i_1}}  \mu'_2
  \fire{w_1} \mu_2$. Since $\mu'_1  \fire{t_{i_1}}  \mu'_2 $, \(Z(t_{i_1})=S_{\ell}\) and
  $S_{\ell}\bigl(O(t_{i_1})\bigr)=\varnothing$, we have
  $S_{\ell}(\mu'_1)=S_{\ell}(\mu'_2)=\varnothing$. Hence, we can apply the
  first case to the runs $\mu_1  \fire{w_0}  \mu'_1$  and  $\mu'_2 \fire{w_1}
  \mu_2$, to show there are $\mmap_1, \mmap'_1,\mmap'_2, \mmap_2 \in
  \multiset{S'}$ such that $S(\mmap_i)=\mu_i$, 
  $S(\mmap'_i)=\mu'_i$, 
  \(R_{\ell}(\mmap'_i)=R_{\ell}(\mmap_i)=\varnothing\)
  for \(i\in\set{1,2}\), 
  $\mmap'_1 \in \fire{\mmap_{1}}^{L_\ell}_{N'}$, and $\mmap_2 \in
  \fire{\mmap'_{2}}^{L_\ell}_{N'}$. 
  Moreover \(t_{i_1}\in T_{\ell}|_T\) shows that
  there exist \(u_{i_1}\in\set{p_{\ell+1},\ldots,p_n}^*\) and 
  \(v_{i_1}\in\set{c_{\ell+1},\ldots,c_n}^*\) such that \(u_{i_1}\cdot t_{i_1}\cdot v_{i_1}\in T_{\ell}\).
  Therefore we can pick \(\mmap_1,\mmap'_1,\mmap'_2,\mmap_2\) such that
  in addition to the above constraints we have
  \(\mmap'_1\fire{u_{i_1}t_{i_1}v_{i_1}}_{N'}\mmap'_2\)
  which is possible since \(\mu'_1\fire{t_{i_1}}\mu'_2\) and
  \(S(\mmap'_i)=\mu'_i\) for \(i\in\set{1,2}\). Finally the above reasoning
  shows that 
  \(\mmap'_1\in\fire{\mmap_1}^{L_{\ell}}_{N'}\),
  \(\mmap'_2\in \fire{\mmap'_1}^{L_{\ell}}_{N'}\),
  \(\mmap_2\in\fire{\mmap'_2}^{L_{\ell}}_{N'}\), hence that
  \(\mmap_2\in\fire{\mmap_1}^{L_{\ell}}_{N'}\) by definition of \(L_{\ell}\) and we are done
  since $S(\mmap_i)=\mu_i$, $R_{\ell}(\mmap_i)=\varnothing$ for \(i\in\set{1,2}\).\qed
\end{itemize}
\end{proof}

\end{document}

%% file: pre.tex
\section{Preliminaries}\label{sec:prelim}

\subsection{Context-Free Languages}

An \emph{alphabet} \(\Sigma\) is a finite non-empty set of \emph{symbols}.
A \emph{word} $w$ over an alphabet \(\Sigma\) is a finite sequence of symbols of
$\Sigma$ where the empty sequence is denoted $\varepsilon$. 
We write $\Sigma^*$ for the set of words over $\Sigma$.
Let $L\subseteq\Sigma^*$, $L$ defines a \emph{language}.

A \emph{context-free grammar} (\(\cfg)\) $G$ is a tuple
$(\mathcal{X},\Sigma,\prod)$ where $\mathcal{X}$ is a finite non-empty set of \emph{variables}
(\emph{non-terminal letters}), $\Sigma$ is an alphabet of
\emph{terminal letters}, and $\prod \subseteq \big(\mathcal{X}\times (\mathcal{X}^2
\cup \Sigma \cup \{\epsilon\}) \big)$ a finite set of \emph{productions} (the
production $(X,w)$ may also be denoted by $X\rightarrow w$). For every production $p=(X,w) \in \prod$, we use $head(p)$ to denote the variable $X$. Observe that the
form of the productions is restricted, but it has been shown in \cite{LL10} that
every \(\cfg\) can be transformed, in polynomial time, into an equivalent
grammar of this form.

Given two strings $u,v \in (\Sigma \cup \mathcal{X})^*$ we define the relation
$u \Rightarrow v$, if there exists a production $(X, w)\in\prod$ and some words
$y,z \in (\Sigma \cup \mathcal{X})^*$ such that $u=yXz$ and $v=ywz$.  We use
$\Rightarrow^*$ for the reflexive transitive closure of $\Rightarrow$.  
Given $X\in \mathcal{X}$, we define the language $L_G(X)$, or simply \(L(X)\)
when \(G\) is clear form the context, as $\set{w\in\Sigma^*\mid X\Rightarrow^*
w}$. 
A language $L$ is \emph{context-free} (\(\cfl\)) if there exists a \(\cfg\) 
$G=(\mathcal{X},\Sigma,\prod)$ and \(A\in\mathcal{X}\) such that
$L=L_G(A)$.

\subsection{Finite-index Approximation of Context-Free Languages}

Let $k\in\nats$, $G=(\mathcal{X},\Sigma,\prod)$ be a \(\cfg\) and
$A\in\mathcal{X}$.  A derivation from \(A\) given by
$A=\alpha_0\Rightarrow\alpha_1\Rightarrow \dots \Rightarrow \alpha_n$ is 
\(k\)-\emph{index bounded} if for every $i\in\set{0,\dots,n}$ at most $k$ symbols of
$\alpha_i$ are variables.  We denote by $L^{(k)}(A)$ the subset of $L(A)$ such
that for every $w\in L^{(k)}(A)$ there exists a $k$ index bounded derivation
$A\Rightarrow^* w$.  We call \(L^{(k)}(A)\) the \emph{\(k\)-index
approximation} of \(L(A)\) or more generically we say that \(L^{(k)}(A)\) is a
\emph{finite-index approximation} of \(L(A)\).\footnote{Finite-index
approximations were first studied in the 60's.} 

Let us now give some known properties of finite-index approximations.
Clearly \(\lim_{k\rightarrow \infty} L^{(k)}(A)=L(A)\).  Moreover, let \(L\) be
a regular or linear language\footnote{See \cite{HMU06} for definitions.}, then
there exists a \(\cfg\) \(G'\), and a variable \(A'\) of \(G'\) such that
\(L(A')=L=L^{(1)}(A')\).  Also Luker showed in \cite{Luker78} that if \(L(A)\subseteq
L(w_1^* \cdots w_n^*) \) for some \(w_i\in\Sigma^*\), then \(L^{(k)}(A)=L(A)\)
for some \(k\in\nats\).  More recently, \cite{EKL08:icalp,gmm10} showed some
form of completeness for finite-index approximation when commutativity of
concatenation is assumed. It shows that there exists a \(k\in\nats\) such that
\(L(A)\subseteq \Pi(L^{(k)}(A))\) where \(\Pi(L)\) denotes the language
obtained by permuting symbols of \(w\) for every \(w\in L\).  As an
incompleteness result, Salomaa showed in \cite{Salomaa1969} that for the Dyck
language \(L_{D_1^*}\) over 1-pair of parentheses there is no \(\cfg\) \(G'\),
variable \(A'\) of \(G'\) and \(k\in\nats\) such that
\(L^{(k)}(A')=L_{D_1^*}\). 

Inspired by \cite{egkl11-ipl,EKL10:JACM,EKL08:icalp} let us define the
\(\cfg\) $G^{[k]}$ which annotates the variables of \(\mathcal{X}\)
with a positive integer bounding the index of the derivations starting with
that variable. 

\begin{definition}
Let
 \(G^{[k]}=(\mathcal{X}^{[k]},\Sigma,\prod^{[k]})\) be the  context-free grammar defined as follows:  \(\mathcal{X}^{[k]}=\set{X^{[i]}\mid 0\leq i \leq k \land X\in\mathcal{X}}\), and  \(\prod^{[k]}\) is the smallest  set such that:
\begin{itemize}
\item For every $X\rightarrow Y\, Z\in \prod$, \(\prod^{[k]}\)
has the productions $X^{[i]} \rightarrow Y^{[i-1]} Z^{[i]}$ and $ X^{[i]} \rightarrow Y^{[i]} Z^{[i-1]}$ for every
$i\in\set{1,\dots,k}$.
\item For every $X \rightarrow \sigma\in\prod$ with  \(\sigma\in\Sigma \cup \{\epsilon\} \), 
$X^{[i]} \rightarrow
\sigma \in \prod^{[k]}$ for all
$i\in\set{0,\dots,k}$.
\end{itemize}
\label{def:cfgbounded}
\end{definition}

What follows is a consequence of several results from different papers by
Esparza \textit{et al}. For the sake of clarity we give a direct proof in the appendix.
{%
\begin{lemma}
	\label{lem1} Let \(X\in\mathcal{X}\). We have \(L(X^{[k]})=L^{(k+1)}(X)\).
\end{lemma}
}

\subsection{Petri nets with Inhibitor Arcs}

Let \(\Sigma\) be a finite non-empty set, a \emph{multiset} $\mmap\colon
\Sigma\mapsto\nats$ over $\Sigma$ maps each symbol of $\Sigma$ to a natural
number.  Let $\multiset{\Sigma}$ be the set of all multiset over $\Sigma$.

We sometimes use the following notation for multisets
$\mmap=\multi{q_1,q_1,q_3}$ for the multiset
$\mmap\in\multiset{\set{q_1,q_2,q_3,q_4}}$ such that $\mmap(q_1)=2$,
$\mmap(q_2)=\mmap(q_4)=0$, and $\mmap(q_3)=1$. The empty multiset is denoted
\(\varnothing\).  

Given $\mmap,\mmap'\in\multiset{\Sigma}$ we define $\mmap\oplus
\mmap'\in\multiset{\Sigma}$ to be the multiset such that $\forall a\in\Sigma\colon
(\mmap\oplus \mmap')(a)=\mmap(a)+\mmap'(a)$, we also define the natural partial order
$\preceq$ on $\multiset{\Sigma}$ as follows: $\mmap\preceq\mmap'$ if{}f there
exists $\mmap^{\Delta}\in\multiset{\Sigma}$ such that
$\mmap\oplus\mmap^{\Delta}=\mmap'$.  We also define
\(\mmap\ominus\mmap'\in\multiset{\Sigma}\) as the multiset such that
\((\mmap\ominus\mmap')\oplus\mmap'=\mmap\) provided \(\mmap'\preceq \mmap\).

A {\em Petri net} with inhibitor arcs ($\pni$ for short)
$N=(S,T,F=\tuple{Z,I,O},\mmap_{\imath})$ consists of a finite non-empty set $S$
of \emph{places}, a finite set $T$ of \emph{transitions} disjoint from $S$, a
tuple $F=\tuple{Z,I,O}$ of functions $Z \colon T \mapsto 2^S$, $I\colon
T\mapsto\multiset{S}$ and $O\colon T\mapsto\multiset{S}$, and an \emph{initial
marking} $\mmap_{\imath}\in\multiset{S}$.  
A marking \(\mmap\) (\(\in \multiset{S}\)) of \(N\) assigns to each place
\(p\in S\) \(\mmap(p)\) \emph{tokens}.

A transition $t\in T$ is \emph{enabled at} $\mmap$, written $\mmap\fire{t}$, if
$I(t)\preceq\mmap$ and $\mmap(p)=0$  for all $p \in Z(t)$.  A transition $t$ that is enabled at $\mmap$ can be 
\emph{fired}, yielding a marking $\mmap'$ such that $\mmap'=(\mmap\ominus I(t))\oplus O(t)$. We write this fact as follows: $\mmap\fire{t}\mmap'$.
We extend enabledness and firing inductively to
finite sequences of transitions as follows. 
Let $w\in T^*$.
If $w=\varepsilon$ we define $\mmap\fire{w}\mmap'$ if{}f 
$\mmap'=\mmap$; else if $w=u\cdot v$ we
have $\mmap\fire{w}\mmap'$ if{}f \(\exists\mmap_{1}\colon\mmap\fire{u}\mmap_{1}\land\mmap_1\fire{v}\mmap'\).  

From the above definition we find that $\mmap$ is a reachable  marking from
$\mmap_{0}$ if and only if there exists $w\in T^*$ such that $\mmap_{0}\fire{w}\mmap$.
Given a language $L\subseteq T^*$ over the transitions of $N$, the \emph{set of
reachable states from $\mmap_{0}$ along $L$}, written $\fire{\mmap_{0}}^L$,
coincides with $\set{\mmap\mid\exists w\in L\colon \mmap_{0}\fire{w}\mmap}$.
Incidentally, if $L$ is unspecified then it is assumed to be $T^*$ and we
simply write $\fire{\mmap_{0}}$ for the set of states reachable from
$\mmap_{0}$. For clarity, we shall sometimes write the \(\pni\) in subscript,
e.g. \(\mmap_1\in\fire{\mmap_0}^L_N\).

A Petri net with {\em weak} inhibitor arcs ($\pnw$ for short)  is a $\pni$ $N=(S,T,F=\tuple{Z,I,O},\mmap_{\imath})$ such that there is an index function \(f\colon S\mapsto\nats\) with the property: %
\begin{equation}
\forall p,p'\in S\colon f(p)\leq f(p') \rightarrow (\forall t\in T\colon p'\in Z(t)\rightarrow p\in Z(t))\enspace .\label{eq:reinhardt}
\end{equation}

A Petri net ($\pn$ for short) can be seen as a subclass of Petri nets with {\em weak} inhibitor arcs where $Z(t)=\emptyset$ for all transitions $t \in T$. In this case, we shorten $F$ as the pair $\tuple{I,O}$.

The {\em reachability} problem for a \(\pni\) $N=(S,T,F=\tuple{Z,I,O},\mmap_{\imath})$ is the problem of deciding, for a given marking  $\mmap$, whether \(\mmap\in\fire{\mmap_{\imath}}\) holds. It is well known that reachability  for Petri nets with inhibitor arcs is  undecidable \cite{hack76}. However, the following holds:

\begin{theorem}{\cite{Reinhardt}}
	The reachability problem for \(\pnw\) is decidable.
	\label{thm:reinhardt}
\end{theorem}

\subsection{The reachability problem for Petri nets along finite-index CFL}

Let us formally define the problem we are interested in.  Given: (1) a Petri net \(N=(S,T,F,\mmap_{\imath})\) where \(T\neq\emptyset\); 
	(2) a \(\cfg\) \(G=(\mathcal{X},T,\prod)\) and \(A\in\mathcal{X}\); (3) a marking \(\mmap_f\in\multiset{S}\);
	and (4) a value \(k\in\nats\).

\noindent
\hspace{0pt}\hspace{\stretch{1}}Does \(\mmap_f\in\fire{\mmap_{\imath}}^{L^{(k)}(A)}\)
hold ?\hfill\vspace{0pt}

%
%

In what follows, we prove the interreducibility of the reachability
problem for \(\pn\) along finite-index \(\cfl\) and the reachability
problem for \(\pnw\).

%% file: pncfl-ipn.tex
%
%
%
\section{From \(\pn\) reachability along \(\text{fi}\cfl\) to \(\pnw\) reachability}\label{sec:reduc_pierre}

In this section, we show that the reachability problem for Petri nets along
finite-index \(\cfl\)  is decidable. To this aim, let us fix an instance of the problem: a Petri net \(N=(S,T,F,\mmap_{\imath})\) where
\(T\neq\emptyset\),  a \(\cfg\) \(G=(\mathcal{X},T,\prod)\),
\(\mmap_f\in\multiset{S}\), and a natural number   \(k\in\nats\). Moreover, let
\(G^{[k]}=(\mathcal{X}^{[k]},T,\prod^{[k]})\) be the \(\cfg\)
given by def.~\ref{def:cfgbounded}. 
	
Lemma \ref{lem1} shows that \(\mmap_f\in\fire{\mmap_{\imath}}^{L^{(k+1)}(A)}\)
if and only if \(\mmap_f\in\fire{\mmap_{\imath}}^{L(A^{[k]})}\). Then, our decision
procedure, which determines if
\(\mmap_f\in\fire{\mmap_{\imath}}^{L(A^{[k]})}\), proceeds by reduction to the
reachability problem for \(\pnw\) and is divided in two steps.  First, we
reduce the question \(\mmap_f\in\fire{\mmap_{\imath}}^{L(A^{[k]})}\) to the
existence of a successful execution in the program of Alg.~\ref{alg:traverse}
which, in turn, is reduced to a reachability problem for \(\pnw\).
Let us describe Alg.~\ref{alg:traverse}.

\noindent %
\begin{minipage}[t]{.33\linewidth}
	{\bf Part 1.} Alg.~\ref{alg:traverse} gives the procedure \(\mathit{traverse}\) in which
 	$\mathbf{M_{\sf i}}$ and ${\mathbf{M_{\sf f}}}$ are
global arrays of markings with index ranging from $0$ to $k$ (i.e., for every
$j \in \{0,\ldots,k\}$, $\mathbf{M_{\sf i}}[j], \mathbf{M_{\sf f}}[j] \in
\mathbb{M}[S]$).  We say that a call \(\mathit{traverse}(X^{[\ell]})\)
\emph{successfully returns} if there exists an execution which eventually
rea\-ches line~\ref{return} (i.e., no assert fails) and the 
postcondition \(\mathbf{M_{\sf i}}[j]={\mathbf{M_{\sf f}}}[j]=
\varnothing\) for every \(j\in\set{0,\ldots,\ell}\) holds. Moreover we say that a
call \(\mathit{traverse}(X^{[\ell]})\) is \emph{proper} if \(\mathbf{M_{\sf
i}}[j]=\mathbf{M_{\sf f}}[j]=\varnothing\) for all \(j<\ell\). 
Let
\(\ell\in\set{0,\ldots,k}\), we shall now demonstrate that a proper call
\(\mathit{traverse}(X^{[\ell]})\) successfully returns if and only if there exists
\(w\in L(X^{[\ell]})\) such that \(\mathbf{M_{\sf
i}}[\ell]\fire{w}_{N}\mathbf{M_{\sf f}}[\ell]\).
\end{minipage}%
\hspace{\stretch{1}}%
\begin{minipage}[t]{.65\linewidth}
\vspace{-3pt}
\begin{algorithm}[H]
	\DontPrintSemicolon
	\LinesNumbered
	\KwIn{A variable $X^{[\ell]}\in\mathcal{X}^{[k]}$ of $G^{[k]}$}
		\Begin{
		Let $p\in\prod^{[k]}$ such that $head(p)=X^{[\ell]}$\;\nllabel{ln:pick}
		\Switch{p}{
		\uCase{\(X^{[\ell]}\rightarrow \sigma\)\tcc*[f]{\(\sigma\in\Sigma \cup \{\epsilon\}\)}}{\nllabel{ln:case0} 
			\(\mathbf{M_{\sf i}}[\ell]:=(\mathbf{M_{\sf i}}[\ell]\ominus I(\sigma))\oplus O(\sigma)\)\; \nllabel{ln:set0}
			\(\subto(\mathbf{M_{\sf i}}[\ell],\mathbf{M_{\sf f}}[\ell])\)\; \nllabel{ln:set1}
		}
		\uCase{\(X^{[\ell]} \rightarrow B^{[\ell]}C^{[\ell-1]}\)}{
			\nllabel{ln:case1}\(\mathit{transfer\_from\_to}({\mathbf{M_{\sf f}}}[\ell],{\mathbf{M_{\sf f}}}[\ell-1])\)\;\nllabel{transfer}
			$\addto(\mathbf{M_{\sf f}}[\ell],\mathbf{M_{\sf i}}[\ell-1])$\;\nllabel{add}
			$\mathit{traverse}(C^{[\ell-1]})$\;\nllabel{traverse_c}
			assert $\mathbf{M_{\sf i}}[j]={\mathbf{M_{\sf f}}}[j]= \varnothing $ \ for all $j<\ell$\;\nllabel{assert1}
			$\mathit{traverse}(B^{[\ell]})$\;\nllabel{traverse_b}
		}
		\Case{\(X^{[\ell]} \rightarrow B^{[\ell-1]}C^{[\ell]}\)}{
		 \nllabel{ln:case2}\(\mathit{transfer\_from\_to}(\mathbf{M_{\sf i}}[\ell],\mathbf{M_{\sf i}}[\ell-1]) \)\;\nllabel{transfer_bis}							 
		 $\addto(\mathbf{M_{\sf i}}[\ell],\mathbf{M_{\sf f}}[\ell-1])$\;\nllabel{add_bis}
		 $\mathit{traverse}(B^{[\ell-1]})$\;  \nllabel{call-l-1_bis}
		 assert $\mathbf{M_{\sf i}}[j]={\mathbf{M_{\sf f}}}[j]= \varnothing $ \  for all $j<\ell$\; \nllabel{assert2}
		 $\mathit{traverse}(C^{[\ell]})$\; \nllabel{traverse_c_bis}
		}
		}
		\Return\;\nllabel{return}
	}
 \caption{$\mathit{traverse}$\label{alg:traverse}}
\end{algorithm}%
\end{minipage}

\noindent %
\hspace*{\stretch{1}}
\begin{minipage}[t]{.5\linewidth}
	\vspace{0pt}
\begin{algorithm}[H]
   {\small    
       \DontPrintSemicolon
			 \KwIn{\(\mathit{src}_1, \mathit{src}_2\)}
       \Begin{
         Let \(\mathit{qty}\) s.t. \(\varnothing \preceq \mathit{qty}\)\;
				 \eIf{\(\addto\)}{%
				 \( (\mathit{src}_1,\mathit{src}_2) := (\mathit{src}_1,\mathit{src}_2)\oplus \mathit{qty}\)\;
				 }(\tcp*[h]{\(\subto\)}){%
				 \( (\mathit{src}_1,\mathit{src}_2) := (\mathit{src}_1,\mathit{src}_2)\ominus \mathit{qty}\)\;
				 }
			 }
			 \caption{\(\addto,\subto\)\label{alg:add}}}
\end{algorithm}%
\end{minipage}%
\hspace{\stretch{1}}
\begin{minipage}[t]{.35\linewidth}
	\vspace{0pt}
\begin{algorithm}[H]
     {\small  
       \DontPrintSemicolon
			 \KwIn{\(\mathit{src}, \mathit{tgt}\)}
       \Begin{
        Let \(\mathit{qty}\) s.t. \(\varnothing \preceq \mathit{qty} \preceq \mathit{src} \)\;
				\(\mathit{tgt} := \mathit{tgt}\oplus \mathit{qty}\)\; 
				\(\mathit{src} := \mathit{src}\ominus \mathit{qty}\)\;
			 }
			 \caption{\(\mathit{transfer\_from\_to}\)\label{alg:transfer}}}
\end{algorithm}%
\end{minipage}%
\hfill\hspace{0pt}

The formal statement is given at Lem.~\ref{lem:alg_equiv}. We give some
intuitions about Alg.~\ref{alg:traverse} first.

\medskip

The control flow of \(\mathit{traverse}\) matches the traversal of a derivation
tree of \(G^{[k]}\) such that at each node \(\mathit{traverse}\) goes first to
the subtree which carries the least index. The tree traversal is implemented
through recursive calls in \(\mathit{traverse}\). To see that the traversal goes first in the subtree
of least index, it suffices to look at the ordering of the recursive calls to
\(\mathit{traverse}\) in the code of Alg.~\ref{alg:traverse}, e.g. in case the
of line~\ref{ln:case1}, \(\mathit{traverse}(C^{[\ell-1]})\) is called before
\(\mathit{traverse}(B^{[\ell]})\).

Reasoning in terms of derivation trees, we have
that the proper call \(\mathit{traverse}(X^{[\ell]})\) returns if{}f there exists a derivation
tree \(t\) of \(G^{[k]}\) with root variable \(X^{[\ell]}\)
such that the sequence of transitions given by the yield of \(t\) is enabled
from the marking stored in \(\mathbf{M_{\sf i}}[\ell]\) and its firing yields the marking
stored in \(\mathbf{M_{\sf f}[\ell]}\).

Because of the least index first traversal, it turns out that the arrays
\(\mathbf{M_{\sf i}}\) and \(\mathbf{M_{\sf f}}\) provide enough space to manage
all the intermediary results.

Also, we observe that when the procedure \(\mathit{traverse}(X^{[\ell]})\)
calls itself with the parameter, say $B^{[\ell]}$, the call is a \emph{tail
recursive call}. This means that when \(\mathit{traverse}(B^{[\ell]})\) returns
then \(\mathit{traverse}(X^{[\ell]})\) immediately returns.  It is known from
programming techniques how to implement tail recursive call without consuming
space on the call stack. In the case of Alg.~\ref{alg:traverse}, we can do so
by having a global variable to store the parameter of \(\mathit{traverse}\) and
by replacing tail recursive calls with \(\mathbf{goto}\) statements.  For the
remaining recursive calls (line \ref{traverse_c} and \ref{call-l-1_bis}),
because the index of the callee is one less than the index of the caller, we conclude that a bounded
space consisting of \(k\) frames suffices for the call stack.

Those two insights (two arrays with \(k\) entries and a
stack with \(k\) frames) will be the key to show, in Part 2, that
\(\mathit{traverse}\) can be implemented as a \(\pnw\).

\begin{lemma}
	Let $\ell \in \{0,\ldots,k\}$, \(X^{[\ell]}\in \mathcal{X}^{[k]}\), and
	$\mmap,\mmap'\in\multiset{S}$. Then, the proper call
	\(\mathit{traverse}(X^{[\ell]})\) with context $\mathbf{M_{\sf i}}[\ell]=\mmap$
  and $\mathbf{M_{\sf f}}[\ell]=\mmap'$
	successfully returns  if and only if  there exists $w \in
	L(X^{[\ell]})$ such that \(\mmap
	\fire{w}_N\mmap'\).\label{lem:alg_equiv}
\end{lemma}
\begin{proof}
{\bf If.}
We prove that if there exists \(w\in L(X^{[\ell]})\) such that
\(\mmap\fire{w}\mmap'\) then the proper call
\(\mathit{traverse}(X^{[\ell]})\) with \(\mathbf{M_{\sf i}}[\ell]=\mmap\) and
\(\mathbf{M_{\sf f}}[\ell]=\mmap'\) successfully returns.

Our proof is done by induction on the length \(n\) of the derivation of \(w\in L(X^{[\ell]})\).
For the case \(n=1\), we necessarily have \(X^{[\ell]}\Rightarrow w=\sigma\) for some \( (X^{[\ell]},\sigma)\in\prod^{[k]}\).
In this case, the proper call \(\mathit{traverse}(X^{[\ell]})\) with \(\mathbf{M_{\sf i}}[\ell]=\mmap\) and
\(\mathbf{M_{\sf f}}[\ell]=\mmap'\) executes as follows: 
\(p=(X^{[\ell]},\sigma)\) is picked and the case of line~\ref{ln:case0} executes successfully
since \(\mmap=\mathbf{M_{\sf i}}[\ell]\fire{\sigma}\mathbf{M_{\sf f}}[\ell]=\mmap'\) holds.
In fact, after the assignment of line \ref{ln:set0} we have \(\mathbf{M_{\sf
i}}[\ell]=\mathbf{M_{\sf f}}[\ell]\). From there, the call to \(\subto\)
can return with \(\mathbf{M_{\sf i}}[\ell]=\mathbf{M_{\sf
f}}[\ell]=\varnothing\) which shows that \(\mathit{traverse}(X^{[\ell]})\)
successfully returns.

\medskip %
For the case \(n>1\), we have \(X^{[\ell]}\Rightarrow^{n} w\) which
necessarily has the form \(X^{[\ell]}\Rightarrow B^{[\ell]} C^{[\ell-1]}\Rightarrow^{n-1} w\)
or \(X^{[\ell]}\Rightarrow B^{[\ell-1]} C^{[\ell]}\Rightarrow^{n-1} w\) by def.\ of \(G^{[k]}\).
Assume we are in the latter case. Thus there exists \(w_1\) and \(w_2\) such
that \(X^{[\ell]}\Rightarrow B^{[\ell-1]} C^{[\ell]}\Rightarrow^{i} w_1 C^{[\ell]}\Rightarrow^{j} w_1 w_2=w\) with \(i+j=n-1\)
and \(\exists \mmap_1\colon \mmap\fire{w_1} \mmap_1\fire{w_2} \mmap'\).
Observe that \(w_1\in L(B^{[\ell-1]})\) and \(w_2\in L(C^{[\ell]})\) and so
by induction hypothesis we find that the proper call \(\mathit{traverse}(B^{[\ell-1]})\)
with \(\mathbf{M_{\sf i}}[\ell-1]=\mmap\), \(\mathbf{M_{\sf f}}[\ell-1]=\mmap_1\) successfully returns.
And so does, by induction hypothesis, the proper call \(\mathit{traverse}(C^{[\ell]})\) with
\(\mathbf{M_{\sf i}}[\ell]=\mmap_1\), \(\mathbf{M_{\sf f}}[\ell]=\mmap'\).
Therefore let us consider the proper call \(\mathit{traverse}(X^{[\ell]})\) with \(\mathbf{M_{\sf i}}[\ell]=\mmap\), \(\mathbf{M_{\sf f}}[\ell]=\mmap'\). We show it successfully returns.

First observe that the call to the procedure \(\mathit{traverse}(X^{[\ell]})\) is proper.
Next, at line~\ref{ln:pick}, pick \(p=(X^{[\ell]},B^{[\ell-1]}C^{[\ell]})\).
Then the call \(\mathit{transfer\_from\_to}(\mathbf{M_{\sf i}}[\ell],\mathbf{M_{\sf i}}[\ell-1])\) of line~\ref{transfer_bis} executes
such that \(\mathbf{M_{\sf i}}[\ell]\) is updated to \(\varnothing\) and \(\mathbf{M_{\sf i}}[\ell-1]\) to \(\mmap\).
Next the call to the procedure \(\mathit{add\_\mathord{\ast}\_to}(\mathbf{M_{\sf i}}[\ell],\mathbf{M_{\sf f}}[\ell-1])\) of line~\ref{add_bis} executes such that
both \(\mathbf{M_{\sf i}}[\ell]\) and \(\mathbf{M_{\sf f}}[\ell-1]\) are updated to \(\mmap_1\). Recall that \(\mmap\fire{w_1}\mmap_1\fire{w_2}\mmap'\).

Finally we showed above that the proper call \(\mathit{traverse}(B^{[\ell-1]})\) successfully returns, the assert that follows too and
finally the proper call \(\mathit{traverse}(C^{[\ell]})\). Moreover it is routine to check that 
upon completion of \(\mathit{traverse}(C^{[\ell]})\) (and therefore
\(\mathit{traverse}(X^{[\ell]})\)) we have \(\mathbf{M_{\sf
i}}[j]=\mathbf{M_{\sf f}}[j]=\varnothing\) for all \(j\leq \ell\).

The left case (i.e. \(p=(X^{[\ell]},B^{[\ell]}C^{[\ell-1]})\in\prod^{[k]}\)) is treated similarly.

\noindent %
{\bf Only If.}
Here we prove that if the proper call \(\mathit{traverse}(X^{[\ell]})\) successfully returns then
there exists \(w\in L(X^{[\ell]})\) such that \(\mathbf{M_{\sf i}}[\ell]\fire{w}_N\mathbf{M_{\sf f}}[\ell]\).

Our proof is done by induction on the number \(n\) of times line~\ref{ln:pick}
is executed during the execution of \(\mathit{traverse}(X^{[\ell]})\). 
In every case, line~\ref{ln:pick} is executed at least once.  
For the case \(n=1\), the algorithm necessarily executes the case of line~\ref{ln:case0}.
The definition of \(G^{[k]}\) shows that along a successful execution of
\(\mathit{traverse}(X^{[\ell]})\), the non deterministic choice of
line~\ref{ln:pick} necessarily returns a production of the form \(
p=(X^{[\ell]},\sigma)\in\prod^{[k]}\). Therefore, a successful execution must
execute line~\ref{ln:set0} and \ref{ln:set1} and then~\ref{return} after
which the postcondition \(\mathbf{M_{\sf i}}[j]=\mathbf{M_{\sf
f}}[j]=\varnothing\) for all \(j\leq \ell\) holds.  Because the postcondition
holds, we find that \(\mathbf{M_{\sf i}}[\ell]=\mathbf{M_{\sf f}}[\ell]\) holds
before executing line~\ref{ln:set1}, hence that \(\mathbf{M_{\sf
f}}[\ell]=\mathbf{M_{\sf i}}[\ell]\ominus I(\sigma)\oplus O(\sigma)\) before
executing line~\ref{ln:set0}, and finally that \(\mathbf{M_{\sf
i}}[\ell]\fire{\sigma}\mathbf{M_{\sf f}}[\ell]\) by semantics of transition \(\sigma\) and we are done.

For the case \(n>1\), the first non deterministic choice of line~\ref{ln:pick}
necessarily picks \(p\in\prod^{[k]}\) of the form \(
(X^{[\ell]},B^{[\ell]}C^{[\ell-1]})\) or \(
(X^{[\ell]},B^{[\ell-1]}C^{[\ell]})\). Let us assume 
\(p=(X^{[\ell]},B^{[\ell]}C^{[\ell-1]})\), hence that the case of
line~\ref{ln:case1} is executed.  Let \(\mmap\) and \(\mmap'\) be respectively
the values of \(\mathbf{M_{\sf i}}[\ell]\) and \(\mathbf{M_{\sf f}}[\ell]\)
when \(\mathit{traverse}(X^{[\ell]})\) is invoked. Now, let
\(\mmap_3,\mmap_{\Delta}\) be such that \(\mmap'=\mmap_3\oplus\mmap_{\Delta}\)
and such that upon completion of the call to \(\mathit{transfer\_from\_to}\) at
line~\ref{transfer} we have that \(\mathbf{M_{\sf f}}[\ell]=\mmap_{\Delta}\)
and \(\mathbf{M_{\sf f}}[\ell-1]=\mmap_3\).  Moreover, let \(\mmap_2\) be the
marking such that \(\mathbf{M_{\sf i}}[\ell-1]=\mmap_2\) upon completion of the
call to \(\addto\) at line~\ref{add}. Therefore we find that
\(\mathbf{M_{\sf f}}[\ell]\) is updated to \(\mmap_{\Delta}\oplus\mmap_2\).
Next consider the successful proper call \(\mathit{traverse}(C^{[\ell-1]})\) of line~\ref{traverse_c} with
\({\mathbf{M_{\sf i}}[\ell-1]}=\mmap_2\), \({\mathbf{M_{\sf f}}[\ell-1]}=\mmap_3\).
Observe that because the execution of \(\mathit{traverse}(X^{[\ell]})\) yields the calls
\(\mathit{traverse}(C^{[\ell-1]})\) and \(\mathit{traverse}(B^{[\ell]})\), we find that the
number of times line~\ref{ln:pick} is executed in \(\mathit{traverse}(C^{[\ell-1]})\) and \(\mathit{traverse}(B^{[\ell]})\)
is strictly less than \(n\).
Therefore, the induction hypothesis shows that there exists \(w_2\) such that
\(w_2\in L(C^{[\ell-1]})\) and \(\mmap_2\fire{w_2}\mmap_3\).  Then
comes the successful assert of line~\ref{assert1} followed by the successful
proper call \(\mathit{traverse}(B^{[\ell]})\) of line~\ref{traverse_b} with
\({\mathbf{M_{\sf i}}[\ell]}=\mmap\) and
\({\mathbf{M_{\sf f}}[\ell]}=\mmap_{\Delta}\oplus\mmap_2\).
Again by induction hypothesis, there exists \(w_1\) such that
\(w_1\in L(B^{[\ell]})\) and \(\mmap\fire{w_1} (\mmap_{\Delta}\oplus \mmap_2)\).

Next we conclude from the monotonicity property of \(\pn\) that since
\(\mmap_2\fire{w_2}\mmap_3\) then
\( (\mmap_2\oplus\mmap_{\Delta})\fire{w_2}(\mmap_3\oplus\mmap_{\Delta})\), hence
that
\(\mmap\fire{w_1}(\mmap_2\oplus\mmap_{\Delta})\fire{w_2}(\mmap_3\oplus\mmap_{\Delta})\)
and finally that \(\mmap\fire{w_1\, w_2}\mmap'\) because \(\mmap'=\mmap_3\oplus\mmap_{\Delta}\).  Finally since
\(w_1w_2\in L(X^{[\ell]})\) we
conclude that \(\mmap'\in\fire{\mmap}^{L(X^{[\ell]})}\) and we are
done.

The left case (i.e. \(p=(X^{[\ell]},B^{[\ell-1]}C^{[\ell]})\in\prod^{[k]}\)) is treated similarly.\qed
\end{proof}
\pagebreak
%

%% file: part2nprg.tex
%
%

\noindent %
{\bf Part 2.} In this section, we  show that it is possible to construct a $\pni$ $N'$ such
that  the problem asking if the call to
\(\mathit{traverse}(A^{[k]})\) successfully returns  can be reduced, in
polynomial time, to a reachability problem for $N'$.
Incidentally, we show that \(N'\) is a \(\pnw\), hence that the reachability problem for $\pn$ along
finite-index $\cfl$ is decidable.

To describe \(N'\) we use a generalization of the net program formalism introduced
by Esparza in \cite{esparza-course} which enrich the instruction set with the test for 0 of a variable.

A \emph{net program} is a finite sequence of \emph{labelled commands} separated by
semicolons.  Basic commands have the following form, where
\(\ell,\ell',\ell_1,\ldots,\ell_k\) are \emph{labels} taken from some arbitrary set,
and \(x\) is a variable over the natural numbers, also called a \emph{counter}.

\medskip

\begin{minipage}[t]{.25\textwidth}
	\begin{itemize}
		\item[] \(\ell\colon x:=x-1\) 
		\item[] \(\ell\colon x:=x+1\) 
		\item[] \(\ell\colon \mathbf{goto} \,\ell'\)
	\end{itemize}
\end{minipage}%
\hspace{\stretch{1}}%
\begin{minipage}[t]{.35\textwidth}
	\begin{itemize}
		\item[] \(\ell\colon \mathbf{if}\, x=0 \, \mathbf{then\ goto}\, \ell' \) 
		\item[] \(\ell\colon \mathbf{goto} \,\ell_1\, \mathbf{or}\cdots \mathbf{or\ goto}\, \ell_k\)
		\item[] \(\ell\colon \mathbf{gosub}\, \ell'\)
	\end{itemize}
\end{minipage}%
\hfill %
\begin{minipage}[t]{.20\textwidth}
	\begin{itemize}
		\item[] \(\ell\colon \mathbf{return}\)
		\item[] \(\ell\colon \mathbf{halt}\)
	\end{itemize}
\end{minipage}

\medskip

A net program is \emph{syntactically correct} if the labels of commands are
pairwise different, and if the destinations of jumps corresponds to existing
labels. Moreover we require the net program to be decomposable into a main
program that only calls first-level \emph{subroutines}, which in turn only call
second level \emph{subroutines}, etc and the jump commands in a subroutine can
only have commands of the same subroutine as destinations.\footnote{Here we
consider the main program as a zero-level subroutine, i.e. jump commands in the
main program can only have commands of the main program as destinations.} Each
subroutine has a unique \emph{entry command} labelled with a subroutine name,
and a unique \emph{exit command} of the form \(\ell\colon\mathbf{return}\).
Entry and exit labelled commands are distinct. 

A net program can only be executed once its variables have received initial
values. In this paper we assume that the initial values are always \(0\).
The semantics of net programs is that suggested by the syntax. 

The compilation of a syntactically correct net program to a \(\pni\) is
straightforward and omitted due to space constraints. See \cite{esparza-course}
for the compilation.

At Alg.~\ref{alg:np} is the net program that implements
Alg.~\ref{alg:traverse}.  In what follows assume \(S\), the set of places of
the underlying Petri net, to be \(\set{1,\ldots,d}\) for \(d\geq 1\).  The
counter variables of the net program are given by \(\{x^{[i]}\}_{0\leq i\leq
k, X\in\mathcal{X}}\) and \(\mathbf{M}_{\sf f}[0..k][1..d]\) \(\mathbf{M}_{\sf
i}[0..k][1..d]\) which arranges counters into two matrices of dimension \(
(k+1)\times d\).  For clarity, our net programs use some abbreviations whose
semantics is clear from the syntax, e.g.  \(\mathbf{M}_{\sf
i}[\ell]:=\mathbf{M}_{\sf i}[\ell]\oplus\mmap\) stands for the sequence
\(\mathbf{M}_{\sf i}[\ell][1]:=\mathbf{M}_{\sf i}[\ell][1]+\mmap(1);
[\ldots];\) \( \mathbf{M}_{\sf i}[\ell][d]:=\mathbf{M}_{\sf
i}[\ell][d]+\mmap(d).\)

Let us now make a few observations of Alg.~\ref{alg:np}:

\noindent %
	\textbullet{} at the top level we have the subroutine \(\mathbf{main}\) which first sets up
	\(\mathbf{M}_{\sf i}[\ell]\) and \(\mathbf{M}_{\sf f}[\ell]\),
	then simulates the call \(\mathit{traverse}(X^{[\ell]})\) and finally checks
	that the postcondition holds (label \(\boldsymbol{0}_{1}\)) before halting (label \(\mathbf{success}\)). 

\noindent %
	\textbullet{} the counter variables \(\{x^{[i]}\}_{0\leq i\leq k, X\in\mathcal{X}}\) defines
		the parameter of the calls to \(\mathbf{traverse}_j\). For instance, a call
		to \(\mathit{traverse}(X^{[j]})\) is simulated in the net program by incrementing \(x^{[j]}\)
		and then calling subroutine \(\mathbf{traverse}_{j}\).

\noindent %
\hspace{\stretch{1}}%
\begin{minipage}[t]{.44\linewidth}
	\vspace{0pt}
	{\footnotesize
\begin{algorithm}[H]
	\SetKw{goto}{goto}
	\SetKw{gosub}{gosub}
	\SetKw{Kor}{or}
	\SetKw{halt}{halt}
\nlset{main:}\(\mathbf{M}_{\sf i}[\ell]:=\mathbf{M}_{\sf i}[\ell]\oplus\mmap\)\;
      \(\mathbf{M}_{\sf f}[\ell]:=\mathbf{M}_{\sf f}[\ell]\oplus\mmap'\)\;
			\(x^{[\ell]}:=x^{[\ell]}+1\)\;
			\gosub \(\mathbf{traverse}_\ell\)\;
\nlset{\(\boldsymbol{0}_1\)}\If{\(\mathbf{M}_{\sf i}[0..\ell]=\varnothing=\mathbf{M}_{\sf f}[0..\ell]\)}{\goto \textbf{success}\;}
\nlset{\(\mathbf{traverse}_j\):} \goto \(\mathbf{p}_1\) \Kor \(\cdots\) \Kor\ \goto \(\mathbf{p}_n\)\;
			[\ldots]\;
\nlset{\(\mathbf{p}_{i_0}\):}  \(x^{[j]}:=x^{[j]}-1\)\;
      \(\mathbf{M}_{\sf i}[j]:=\mathbf{M}_{\sf i}[j]\ominus I(\sigma)\)\;
      \(\mathbf{M}_{\sf i}[j]:=\mathbf{M}_{\sf i}[j]\oplus O(\sigma)\)\;
			\gosub \(\mathbf{sub\_to}_{j}\)\;
			\goto \textbf{exit}\;
			[\ldots]\;
\nlset{\(\mathbf{p}_{i_1}\):}  \(x^{[j]}:=x^{[j]}-1\)\;
			\gosub \(\mathbf{tr\_f}_{j}\mathbf{\_f}_{(j-1)}\)\;
			\gosub \(\mathbf{add\_to\_i}_{(j-1)}\_\mathbf{f}_{j}\)\;
      \(c^{[j-1]}:=c^{[j-1]}+1\)\;
			\gosub \(\mathbf{traverse}_{(j-1)}\)\;
			\nlset{\(\boldsymbol{0}_2\)}\If{\(\mathbf{M}_{\sf i}[0..j-1]\!=\!\varnothing\!=\!\mathbf{M}_{\sf f}[0..j-1]\)}{\goto \textbf{l1}\;}
			\nlset{l1:}  \(b^{[j]}:=b^{[j]}+1\)\;
			\goto \(\mathbf{traverse}_j\)\;
			[\ldots]\;
\nlset{\(\mathbf{p}_{i_2}\):} \(v^{[j]}:=v^{[j]}-1\)\;
			\gosub \(\mathbf{tr\_i}_{j}\mathbf{\_i}_{(j-1)}\)\;
			\gosub \(\mathbf{add\_to\_i}_{j}\_\mathbf{f}_{(j-1)}\)\;
			\(y^{[j-1]}:=y^{[j-1]}+1\)\;
			\gosub \(\mathbf{traverse}_{(j-1)}\)\;
			\nlset{\(\boldsymbol{0}_3\)}\If{\(\mathbf{M}_{\sf i}[0..j-1]\!=\!\varnothing\!=\!\mathbf{M}_{\sf f}[0..j-1]\)}{\goto \textbf{l2}\;}
			\nlset{l2:}  \(z^{[j]}:=z^{[j]}+1\)\;
			\goto \(\mathbf{traverse}_j\)\;
			[\ldots]\;
			\nlset{exit:} \Return{}\;
			[\ldots]\;
			\nlset{success:} \halt\;
\caption{ \footnotesize \textbf{main} invoking \(\mathit{traverse}(X^{[\ell]})\) with \(\mmap\),
\(\mmap'\) and subroutines \(\mathbf{traverse}_j\) where \(0<
j\leq\ell\) implementing the calls \(\set{\mathit{traverse}(X^{[j]})}_{X\in\mathcal{X}}\).\label{alg:np}}
\end{algorithm}%
			}
\end{minipage}%
\hspace{\stretch{1}}%
\begin{minipage}[t]{.50\linewidth}
	\noindent %
		\textbullet{} the non-deterministic jump at label \(\mathbf{traverse}_{j}\) simulates
		the selection of a production rule \(\mathbf{p}_{i_k}=(X^{[j]},w)\) which will be fired next
		(if enabled else the program fails). 

	\noindent %
		\textbullet{} 
the missing code for the subroutines
		\(\mathbf{tr\_f}_{j}\mathbf{\_f}_{j-1}\),
		\(\mathbf{add\_to\_i}_{j}\_\mathbf{f}_{j-1}\),
		and \(\mathbf{sub\_to}_{j}\) can be found in the appendix although it is pretty obvious to infer from
		Alg.~\ref{alg:add} and Alg.~\ref{alg:transfer}. The code for \(\mathbf{tr\_i}_{j}\mathbf{\_i}_{j-1}\),
		\(\mathbf{add\_to\_f}_{j}\_\mathbf{i}_{j-1}\) and \(\mathbf{traverse}_{0}\) is also routine to write.

		\noindent %
		\textbullet{}
the program is syntactically correct. First, the levels are assigned to
		subroutines as follows: the level of \(\mathbf{traverse}_j\) is \(j\), the
		level of \(\mathbf{tr\_f}_{j}\mathbf{\_f}_{j-1}\),
		\(\mathbf{tr\_i}_{j}\mathbf{\_i}_{j-1}\),
		\(\mathbf{add\_to\_i}_{j}\_\mathbf{f}_{j-1}\),
		\(\mathbf{add\_to\_f}_{j}\_\mathbf{i}_{j-1}\) and \(\mathbf{sub\_to}_{j}\)
		is \(j-1\).  Given that level assignment, it is routine to check that
		subroutines of level \(i\) only call subroutines of level \(i-1\).
		Moreover, thanks to the programming techniques that allow to implement the
		tail recursive call as a \textbf{goto} instead of \textbf{gosub} we find
		that the program is synctactically correct.  (If we had used \textbf{gosub}
		everywhere, then the net program would be synctactically incorrect).  Also
		observe that each jump commands does not leave the subroutine inside which it is
		invoked.  

\noindent %
\textbullet{} the tests for 0 (labels
		\(\boldsymbol{0}_1,\boldsymbol{0}_2,\boldsymbol{0}_3\)) have a particular
		structure matching the level of the subroutines (level \(0\) for
		\(\boldsymbol{0}_1\) and \(j\) for \(\boldsymbol{0}_2\) and
		\(\boldsymbol{0}_3\)). So, after compilation of the net program into a \(\pni\) \(N'\),
		if we set a mapping \(f\) from the places of
		\(N'\) to \(\nats\) such that \(c\) is mapped to \(i\) if \(c\in
		\set{\mathbf{M}_{\sf i}[i][j]\mid
		j\in\set{1,\ldots,d}}\cup\set{\mathbf{M}_{\sf f}[i][j]\mid
		j\in\set{1,\ldots,d}}\) and every other place is mapped to \(\ell+2\) then we
		find that \(N'\) is a \(\pnw\). Clearly, deciding whether
		Alg.~\ref{alg:np} halts reduces to \(\pnw\) reachability.
		Therefore, by Thm.~\ref{thm:reinhardt}, it is decidable whether Alg.~\ref{alg:np} halts.
\end{minipage}%
\hfill

\begin{lemma}
\label{lem2-part2}
Let $\ell \in \set{0,\ldots,k}$, $X^{[\ell]}\in\mathcal{X}^{[k]}$, and
$\mmap,\mmap'\in\multiset{S}$. Then the proper call \(\mathit{traverse}(X^{[\ell]})\) with 
$\mathbf{M_{\sf i}}[\ell]=\mmap$, $\mathbf{M_{\sf f}}[\ell]=\mmap'$
successfully returns if{}f Alg.~\ref{alg:np} halts.
\end{lemma}
Hence from Lem.~\ref{lem1}, \ref{lem:alg_equiv} and \ref{lem2-part2}, we conclude the following.
\begin{corollary}
The reachability problem for
\(\pn\) along finite-index \(\cfl\)  can be reduced to the reachability problem for \(\pnw\). 
\end{corollary}

%% file: pnw-pncfl.tex
\section{From $\pnw$ reachability to  $\pn$ reachability along \(\text{fi}\cfl\)}\label{sec:reduc_faouzi}

In this section, we show that the reachability problem for \(\pnw\) can be
reduced to the reachability problem of \(\pn\) along finite-index \(\cfl\). To
this aim, let $N=(S,T,F=\tuple{Z,I,O},\mmap_{\imath})$ be a $\pnw$, $\mmap_f
\in \multiset{S}$ a marking,  and \(f\colon S\mapsto\nats\) an
index function such that \eqref{eq:reinhardt} holds.


Let $S=\{s_1,\ldots,s_{n+1}\}$   and $T=\{t_1,\ldots, t_m\}$.  Because it
simplifies the presentation we will make a few assumptions that yield no loss
of generality.  (\(i\)) For every $i \in \set{1,\ldots,n}$, we have $f(s_i) \leq
f(s_{i+1})$, (\(ii\)) $\mmap_{\imath}=\multi{s_{n+1}}$,   $\mmap_{f}=\varnothing$,
(\(iii\)) $Z(t_1) \subseteq Z(t_2) \subseteq \cdots \subseteq Z(t_m) \subseteq
\{s_1,\ldots,s_n\}$, and (\(iv\)) for every $t \in T$, if $s \in Z(t)$ then
$O(t)(s)=0$ (see \cite{Reinhardt}, Lemma 2.1).  Notice that the Petri net $N$
can not test if the place $s_{n+1}$ is empty or not.

In the following, we show that it is possible to construct a Petri net (without
inhibitor arcs)  $N'$, a marking
\(\mmap'_{f}\), and a finite-index \(\cfl\) $L$ such that:
\(\mmap_{f}\in\fire{\mmap_{\imath}}_{N}^{T^*}\)
if{}f
\(\mmap'_{f}\in\fire{\mmap'_{\imath}}_{N'}^{L}\).

\smallskip %
\noindent %
{\bf Constructing  the Petri net $N'$:} 
Let \(N'=(S',T',F'=\tuple{I',O'},\mmap'_{\imath})\) be a \(\pn\) which consists
in \(n+1\) unconnected \(\pn\) widget:
the widget \(N_0\) given by \(N\) without tests for zero (i.e. \(Z(t)\) is set to \(\varnothing\) for every \(t\in T\)) and the widgets \(N_1,\ldots,N_n\) where each
\(N_{i}=(\set{r_i},\set{p_i,c_i}, F_i, \varnothing)\) where
\(F_i(p_i)=\tuple{\varnothing,\multi{r_i}}\) and
\(F_i(c_i)=\tuple{\multi{r_i},\varnothing}\). 
{\(N_i\) is depicted as follows:
\(\stackrel{p_i}{\blacksquare}\!\mathord{\rightarrow}\!\mathord{\stackrel{r_i}{\bigcirc}}\mathord{\rightarrow}\stackrel{c_i}{\blacksquare}\).}
Finally, define \(\mmap'_{\imath}\in\multiset{S'}\) to be 
$\mmap'_{\imath}(s)=\mmap_{\imath}(s)$ for $s \in S$ and $0$ elsewhere;
and \(\mmap'_f=\varnothing\).

{
Since we have the ability to restrict the possible sequences of transitions
that fire in \(N'\), we can enforce the invariant that the sum of tokens in
\(s_i\) and \(r_i\) stays constant.  To do so it suffices to force that
whenever a token produced in \(s_i\) then a token is consumed from \(r_i\) and
vice versa. Call \(L\) the language enforcing that invariant. Then, let
\(\mmap\) be a marking such that \(\mmap(s_i)=\mmap(r_i)=0\), observe that by
firing from \(\mmap\) a sequence of the form: (\(i\)) \(p_i\) repeated \(n\) times, (\(ii\))
any sequence \(w\in L\) and (\(iii\)) \(c_i\) repeated \(n\) times; the marking
\(\mmap'\) that is reached is such that \(\mmap'(s_i)=\mmap'(r_i)=0\). This
suggests that to simulate faithfully a transition \(t_0\) of \(N\) that does
test \(s_i\) for \(0\) we allow the occurrence of the counterpart of \(t_0\) in
\(N_0\) right before (\(i\)) or right after (\(iii\)) only. In what follows, we build
upon the above idea the language \(L_n\) which, as we we will show, coincides
with the finite-index approximation of some \(\cfg\).
}

We need the following notation. Given  a word $v \in \Sigma^*$ and \(\Theta\subseteq
\Sigma\), we define $v|_{\Theta}$ to be the word obtained from $v$ by erasing
all the symbols that are not in $\Theta$. We extend it to languages as follows:
Let \(L\subseteq\Sigma^*\). Then $L|_{\Theta}=\set{u|_{\Theta}\,\mid \, u \in
L}$.

\smallskip %
\noindent %
{\bf Constructing the language \(L_n\):} For every $j  \in \{1,\ldots,m\}$,
let ${u}_j= p^{\, i_1}_1 p^{\, i_2}_2 \cdots p^{\, i_n}_n$ and  $v_j= c^{\,
k_1}_1 c^{\, k_2}_2 \cdots c^{\, k_n}_n$  be two words over  the alphabet $T'$
such that $i_\ell=I(t_j)(s_\ell)$ and $k_\ell=O(t_j)(s_\ell)$ for all $\ell \in
\{1,\ldots,n\}$.   Observe that  firing $v_j
t_j u_j$ keeps unchanged the total number of tokens in \(\set{s_i,r_i}\) for
each \(i\in\set{1,\ldots,n}\). Let
\(\ell\in\set{0,\ldots,n}\) define \(T_{\ell}=\set{ v_j\cdot t_j\cdot u_j \mid
Z(t_j)=\set{s_1,\ldots,s_{\ell}}}\).\footnote{Note that if \(\ell=0\) then
\(\set{s_1,\ldots,s_{\ell}}=\emptyset\).} Also given \(a,b\in\Sigma^*\) and
\(Z\subseteq \Sigma^*\), define \( \tuple{a,b}\star Z\) as the set \(\set{ a^i \cdot
z \cdot b^i \mid i\in\nats\land z\in Z}\).

Define the \(\cfl\)s \(L_0,\ldots,L_n\) inductively as follows: \(L_0=T_{0}^*\)
and for \(0<\ell\leq n\) define \(L_\ell= \bigl((\tuple{p_{\ell},c_{\ell}}\star
L_{\ell-1}) \cup T_{\ell}\bigr)^*\).  It is routine to check that $L_0
\subseteq L_1 \subseteq \cdots \subseteq L_n$ (since $L_{\ell-1}\subseteq
\tuple{p_{\ell},c_{\ell}}\star L_{\ell-1})$) and $L_n|_{T}=T^*$ (since
\(L_n\supseteq \bigcup_{i=0}^n T_i\)). 
{%
Also, \(L_0\) is a regular
language and therefore there exists a \(\cfg\) \(G_0\) and a variable \(A_0\)
of \(G_0\) such that \(L^{(1)}(A_0)=L_0\).  Now, let us assume that for \(L_i\)
there exists a \(\cfg\) \(G_i\) and a variable \(A_i\) such that
\(L^{(i+1)}(A_i)=L_i\).  From the definition of \(L_{i+1}\) it is routine to
check that there exists a \(\cfg\) \(G_{i+1}\) and a variable \(A_{i+1}\) such
that \(L^{(i+2)}(A_{i+1})=L_{i+1}\). Finally we find that \(L_{n}\) can be
captured by the \(n+1\)-index approximation of a \(\cfg\).
}

\begin{lemma}
\label{sec5.lem7}
Let $\ell \in \{0,\ldots,n\}$. If $\mmap_1, \mmap_2 \in \multiset{S'}$ such
that $\mmap_2 \in \fire{\mmap_{1}}^{L_\ell}_{N'}$, then
$\mmap_2(s_j)+\mmap_2(r_j)=\mmap_1(s_j)+\mmap_1(r_j)$  for all $j \in
\{1,\ldots,n\}$.
\end{lemma}%
Let us make a few observations about the transitions of \(N'\) which were
carrying out 0 test in \(N\). In \(L_{\ell}\) no transition \(t\) such that
\(s_{\ell+1}\in Z(t)\) is allowed, that is no test of place \(s_{\ell+1}\) for
\(0\) is allowed along any word of \(L_{\ell}\). The language \(L_{\ell}\)
imposes that the place \(s_{\ell}\) can only be tested for \(0\) along
\(T_{\ell}\). The intuition is that \(L_{\ell}\) allows to test
\(s_{\ell}\) for \(0\) provided all places $s_j$ and $r_j$ for $j \leq \ell$
are empty.

Let us introduce the following notations. Let \(\mmap\in\multiset{S'}\) and
\(Q\subseteq S'\), we write \(Q(\mmap)\) for the multiset of \(\multiset{Q}\)
such that \(Q(\mmap)(q)=\mmap(q)\) for all \(q\in Q\).  We define the following
subsets of places of \(N'\): \(R_{\ell}\) (resp.  \(S_{\ell}\)) is given by
\(\set{r_1,\ldots,r_{\ell}}\) (resp.  \(\set{s_1,\ldots,s_{\ell}}\)).  The proofs of lemmata that follow
are done by induction and given in the appendix.

\begin{lemma}
\label{sec5.lem8}
Let \(\ell\in\set{0,\ldots,n}\), \(w\in L_{\ell}\), and
\(\mmap_a,\mmap_b\in\multiset{S'}\) such that \linebreak \((S_{\ell}\cup
R_{\ell})(\mmap_a)=(S_{\ell}\cup
R_{\ell})(\mmap_b)=\varnothing\) and $\mmap_{a}\fire{w}_{N'} \mmap_b$. Then   $S(\mmap_a)
\fire{w|_{T}}_{N} S(\mmap_{b})$.
\end{lemma}

\begin{lemma}
\label{sec5.lem9}
Let $\ell \in \{0,\ldots,n\}$, $\mu_1, \mu_2 \in \multiset{S}$ such that  $S_{\ell}(\mu_1)=S_{\ell}(\mu_2)=\varnothing$ and  $\mu_2 \in \fire{\mu_1}^{L_\ell|_T}_N$. Then there are $\mmap_1, \mmap_2  \in \multiset{S'}$ such that $S(\mmap_1)=\mu_1$, $S(\mmap_2)=\mu_2$, $R_{\ell}(\mmap_1)=R_{\ell}(\mmap_2)=\varnothing$, and   $\mmap_2 \in \fire{\mmap_{1}}^{L_\ell}_{N'}$.
\end{lemma}

\begin{lemma}
\label{sec.th1}
\(\mmap_f (=\varnothing) \in \fire{\mmap_{\imath}}_N\) if and only if \(\mmap'_f (=\varnothing) \in \fire{\mmap'_{\imath}}^{L_n}_{N'}\).
\end{lemma}
\begin{proof}
({\em $\Rightarrow$}) Assume that $\mmap_f \in \fire{\mmap_{\imath}}_N$. 
Since $L_n|_T=T^*$ and
$S_{n}(\mmap_{\imath})=S_{n}(\mmap_f)=\varnothing$, the result of Lem.~\ref{sec5.lem9}
shows that there are $\mmap_1, \mmap_2  \in \multiset{S'}$ such that
$S(\mmap_1)=\mmap_{\imath}$, $S(\mmap_2)=\mmap_f$, $R_{n}(\mmap_1)=R_{n}(\mmap_2)=\varnothing$,
and $\mmap_2 \in \fire{\mmap_{1}}^{L_n}_{N'}$. This implies that $\mmap'_f
\in \fire{\mmap'_{\imath}}^{L_n}_{N'}$ since $\mmap'_f=\mmap_2$ and
$\mmap'_{\imath}=\mmap_1$ by definition.

\medskip
\noindent
({\em $\Leftarrow$}) Assume that $\mmap'_f \in
\fire{\mmap'_{\imath}}^{L_n}_{N'}$. 
The definition of \(\mmap'_{\imath}\) and \(\mmap'_f\)  shows that
\( (S_n\cup R_n)(\mmap'_{\imath})=(S_n\cup R_n)(\mmap'_f)=\varnothing \)
and therefore,  by Lem.~\ref{sec5.lem8}, we find 
that $S(\mmap'_f) \in \fire{S(\mmap'_{\imath})}^{L_n|_T}_N$, hence that
\(\mmap_f \in\fire{\mmap_{\imath}}^{L_n|_T}_N\) by definition of \(\mmap_{\imath}, \mmap_f\), and finally that \(\mmap_f \in\fire{\mmap_{\imath}}_N\) since \({L_n|_T}=T^{*}\).
\qed
\end{proof}

As an immediate consequence of Lemma \ref{sec.th1}, we obtain the following
result:

\begin{corollary}
The reachability problem for \(\pnw\) can be reduced to the reachability problem for
\(\pn\) along finite-index \(\cfl\). 
\end{corollary}